\documentclass[10pt]{article}
\usepackage[utf8]{inputenc} 
\usepackage[left=2cm,right=2cm,top=2cm,bottom=2cm]{geometry}
\usepackage{graphicx}
\usepackage{caption}
\usepackage{subcaption}
\usepackage{amsmath}
\usepackage[backend=biber,sortcase=true,sortcites=true,sorting=none,style=science,dateabbrev=true]{biblatex}
\usepackage{indentfirst}
\usepackage{authblk}
\bibliography{main.bib}

\title{Half-metallicity in $Ni_2XMn$ Heusler alloys ($X=Fe,Co,Cr$): ab-initio calculations} 
\author[1,2,*]{Alejandro Alés}
\affil[1]{Instituto de Física de Materiales Tandil (IFIMAT), Universidad Nacional del Centro de la Provincia de Buenos Aires (UNCPBA), Pinto 399, 7000 Tandil, Argentina}
\affil[2]{Consejo Nacional de Investigaciones Científicas y Técnicas, Godoy Cruz 2290, Buenos Aires, Argentina}
\affil[*]{Corresponding author: Alejandro Alés, aales@ifimat.exa.unicen.edu.ar}
 
\begin{document}

\maketitle

\begin{abstract}

Lattice parameter, bulk modulus, formation energy and magnetism equilibrium structure in all-d transition metals $Ni_2XMn$  (X=Co,Cr,Fe) for Full and Inverse Heusler structure are studied by first-principles calculations. Tetragonal distortion is calculated for the equilibrium volume of each alloy and possible new equilibrium structures are reported. Density of electronic states are analyzed and the results lead to half-metallicity for this type of alloys.
\end{abstract}

\section{Introduction}

Full Heusler (Inverse Heusler) are ternary intermetallic alloys with composition $X_2YZ$ in space group No 225 (No 216) and the prototype of $Cu_2MnAl$ ($Hg2_2TiCu$) structure, also namely $L2_1$ ($f\bar{4}3m$) in literature\cite{galanakis2016theory,wollmann2017heusler}. Commonly, $X$ and $Y$ are d-group element and $Z$ is main group element\cite{webster1969heusler,graf2011simple}. This combination leads to a p-d orbital hybridization that conferring mechanical properties not always useful from an mechanical point of view \cite{bachaga2019nimn}. Although they have been studied for decades, in recent times the Heusler alloys have increased their interest for different recently discovered applications, such as spintronics, magnetocaloric effect, which add to shape memory properties \cite{graf2010heusler,jiang2021review,palmstrom2016heusler,tavares2022heusler} 

In recent years, Heusler alloys where all the element are d-transition metals are attracting attention because they present interesting mechanical and magnetic properties\cite{wei2015realization,de2021all}. $Ni-Mn$ based alloys are no exception in this trend; previous work were showed that hybridization of the orbitals of these atoms joint with a p-type element produces magnetic effects of great interest, such as giant elastocaloric in $Ni-Mn-In$ \cite{priolkar2011role,huang2015giant} and $Ni-Mn-Sn$ \cite{alvarez2017conventional}. On other hand, recent reports showed that $Ni-Mn$ in combination with Titanium (d-type element), barocaloric and mechanocaloric \cite{aznar2019giant}, colossal elastocaloric effect was reported by Cong et al \cite{cong2019colossal} and Li et al \cite{li2022dd}. This seems to be associated with an interesting property of the d-orbitals of manganese, whose combination with the external orbitals of the rest of the components has a prominent role in the magnetism of the alloy \cite{plogmann1999local}. It is a matter of discussion whether the magnetic interaction of manganese atoms can be described as Ruderman-Kittel-Kasuya-Yosida (RKKY) theory is sufficient \cite{kubler1983formation} or more complex theories are required \cite{wollmann2014magnetism}.

In this brief article, basic properties for stoichiometric alloys of the Heusler-type $Ni_2XMn$ based with d-transition elements (X=Cr,Co,Fe) is studied by first-principles method. Due to the fact that all atomic species present magnetic properties in their pure state, as nickel, iron and cobalt are ferromagnetic and have a high Curie temperature, chromium and manganese are antiferromagnetic with a Néel temperature close to the temperature ambient, alloys between these elements are expected to give rise to complex magnetic structures of great interest from a technological point of view. In the rest of the paper, we detailed the computational calculation used in Sec. \ref{sc:CM} and the results are presented in Sec. \ref{sc:RA}. The conclusions of this report are showed in Sec. \ref{sc:CC}.

\section{Computational methods}
\label{sc:CM} 

Spin-polarized ab-initio calculations were carried out with open-source Quantum-Espresso suite\cite{qe1,qe2} based in density functional theory (DFT), that employ plane-wave expansion and pseudopotentials for description of core electrons. Generalized gradient approximation (GGA) was parameterized by PBEsol exchange-correlation functional\cite{Pbe1,Pbe2,perdew2008restoring}, by means Ultrasoft pseudopotentials\cite{Vanderbilt}. Cut-off energy for wave functions  were expanded to a cut-off $75$ Ry and the density of electrons had a cut-off energy of $550$ Ry. The integration in the Brillouin zone was realized in a 12x12x12 vector grid of the Monkhorst–Pack type, automatically generated by Quantum-Espresso' subroutine. This configuration of parameters was tested for every atomic specie with satisfactory accuracy of less than $10^{-3}$ meV.

\begin{figure}[ht]
  \begin{center}
    \includegraphics[width=0.75\textwidth]{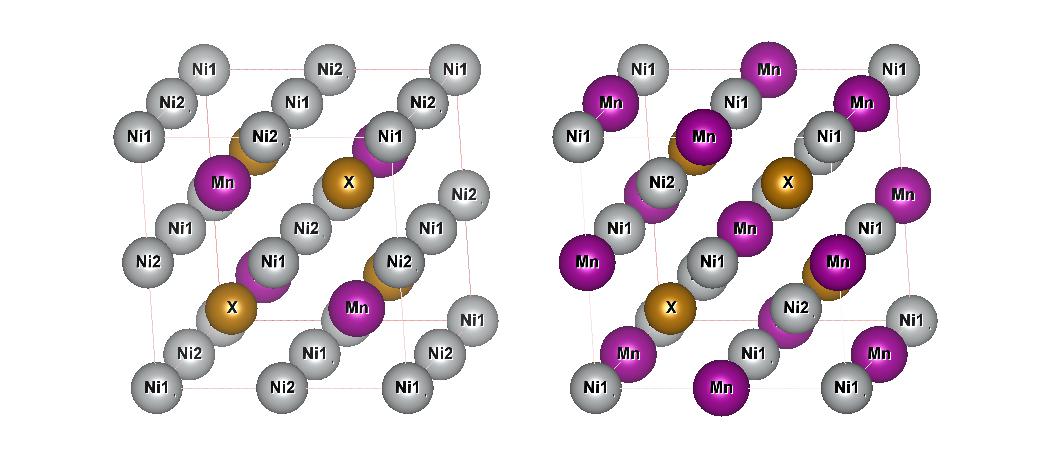}
     \caption{Full Heusler $L2_1$ (left) and Inverse Heusler $f\bar{4}3m$ (right) structures.}
     \label{fig:Estructuras}
 \end{center}
\end{figure}
 
The structures studied are show in Fig. \ref{fig:Estructuras}. Them are derived from the bcc lattice for the composition $X_2 Y Z$ and can be analyzed by means of four interpenetrating fcc sublattices, where each atomic species occupies a sublattice, where grey color are used for nickel atom, violet for manganese atom and X atom are describe for orange color. The Full Heusler alloy with structure $L2_1$ shown in Subfig. \ref{fig:Estructuras}(a) and the Inverse Heusler $f\bar{4}3m$ in Subfig. \ref{fig:Estructuras}(b). We have marked the sublattices for Ni atoms with the index $1,2$ because they could exhibit different magnetic properties for subsequent calculations. 

The equilibrium structure was optimized by calculating the energies for different volumes and the subsequent adjustment by the Birch-Murnaghan equation of state\cite{Murnaghan244}. The initial magnetic configurations have been of the ferromagnetic (FM) and antiferromagnetic (AFM) type. Magnetic moment per atomic specie is obtained as difference of spin-up and spin-down electronic density of states (DOS). Once the equilibrium structure was obtained, we proceeded to calculate the tetragonal distortion and the density of electronic states.

\section{Results and Analysis}
\label{sc:RA} 

\begin{figure*}[tbhp]
\centering
    \begin{subfigure}[t]{0.30\textwidth}
    \centering
     \includegraphics[width=\textwidth]{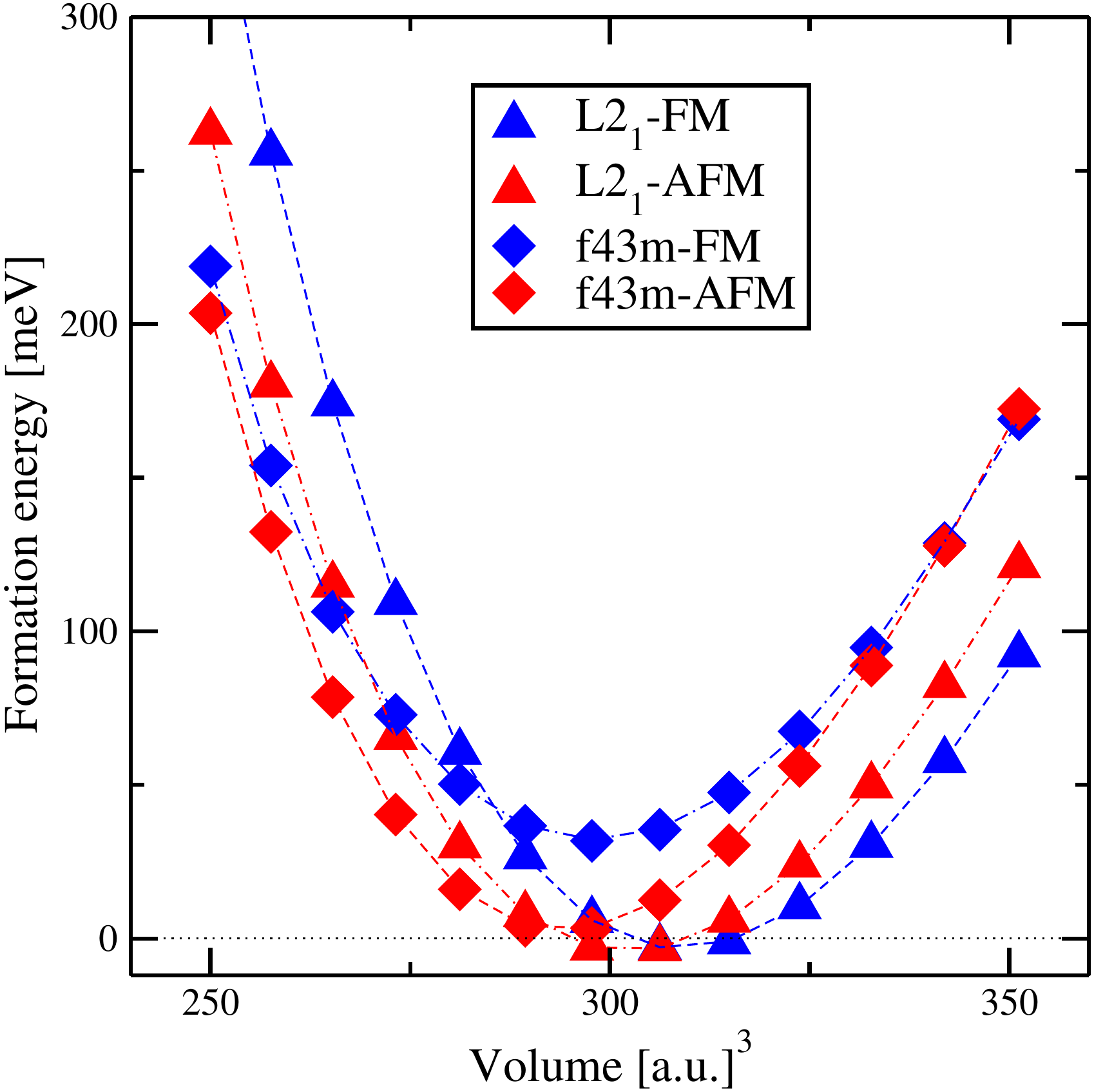}
     \caption{ $Ni_2FeMn$.}
     \label{fig:Ni2FeMn}
 \end{subfigure} \hspace{1em}
\begin{subfigure}[t]{0.30\textwidth}
\centering
     \includegraphics[width=\textwidth]{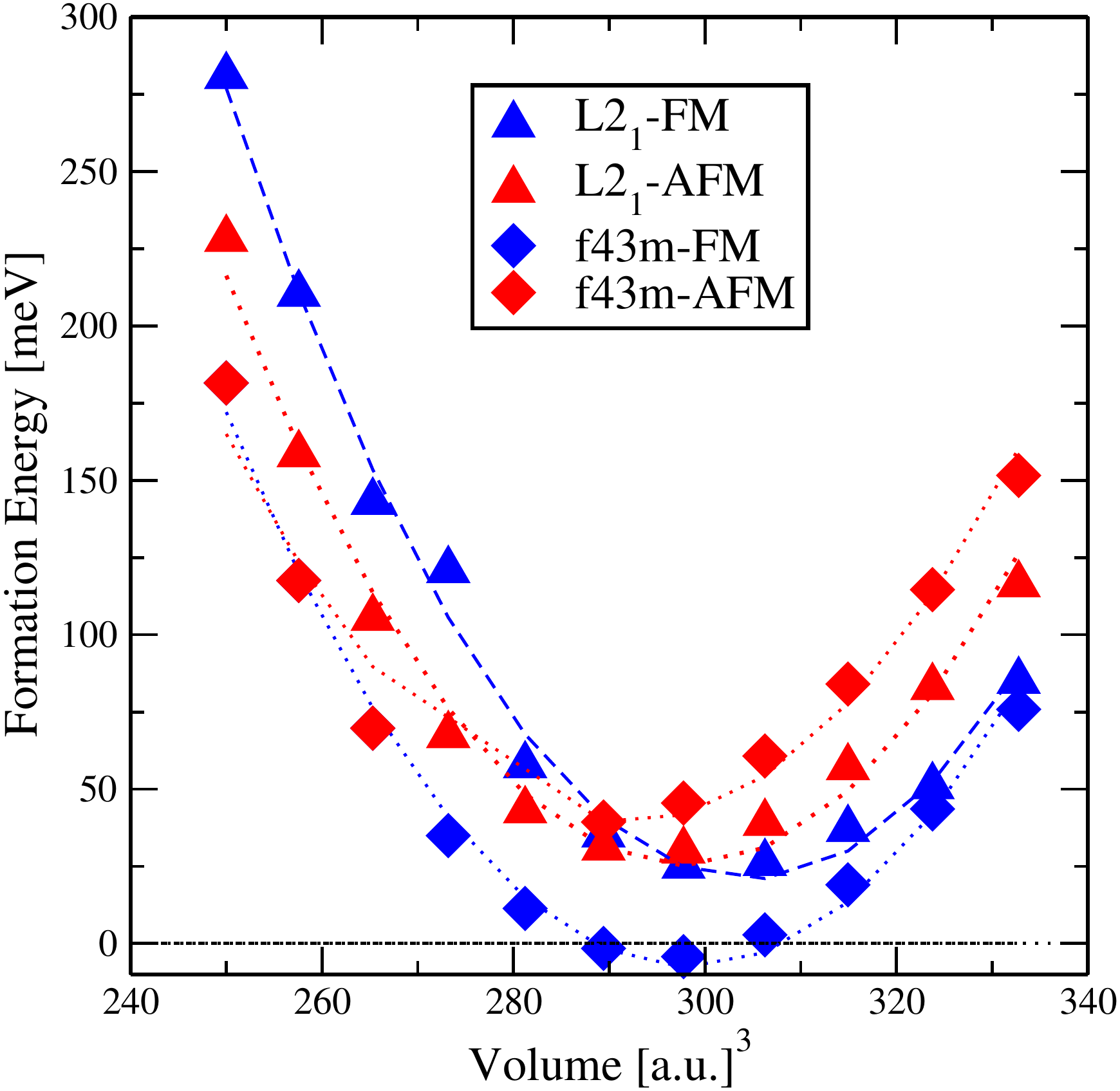}
     \caption{ $Ni_2CoMn$.}
     \label{fig:Ni2CoMn}
 \end{subfigure}\hspace{1em}
\begin{subfigure}[t]{0.30\textwidth}
\centering
     \includegraphics[width=\textwidth]{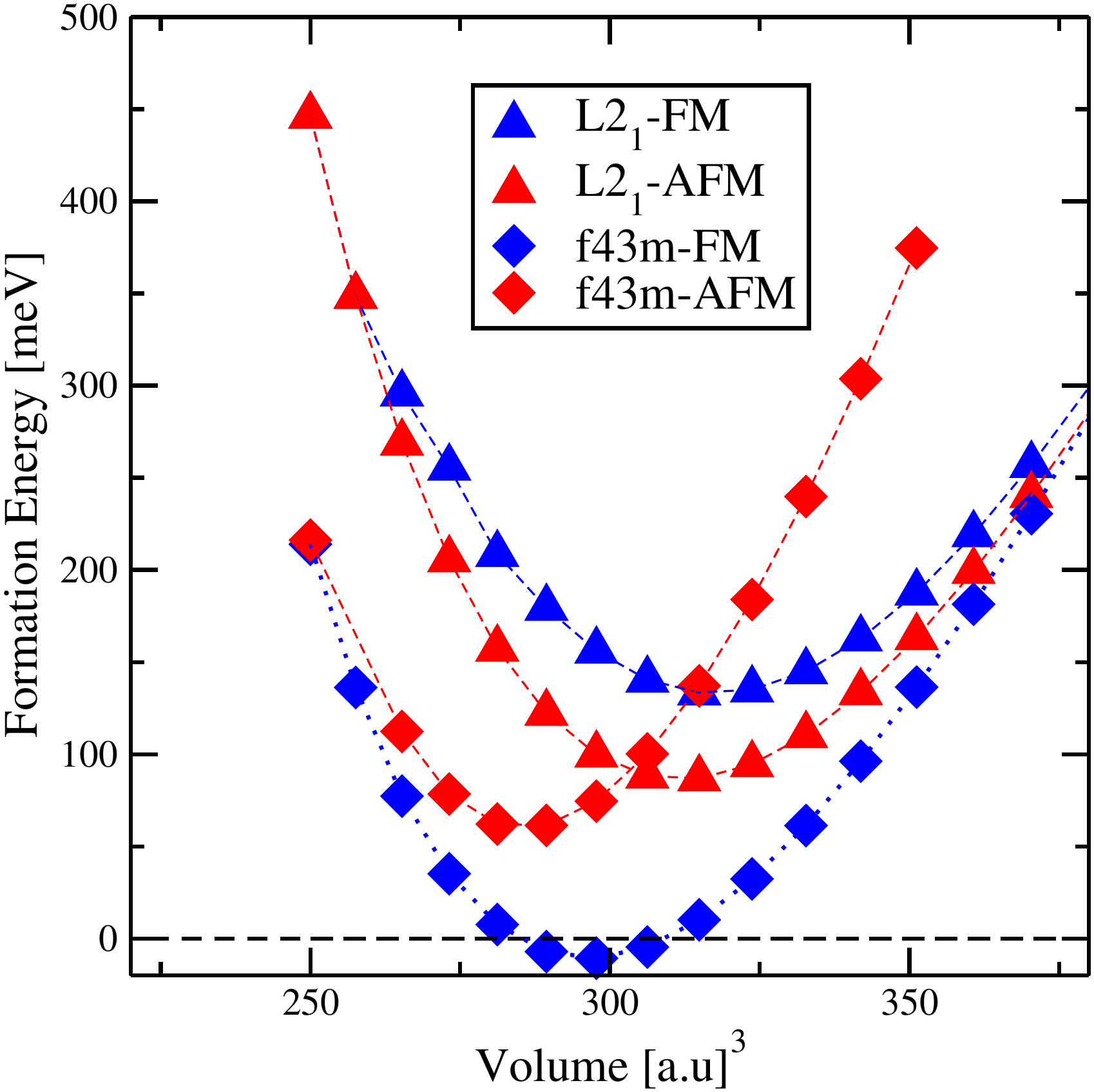}
     \caption{ $Ni_2CrMn$}
     \label{fig:Ni2CrMn}
 \end{subfigure}\hspace{1em}
 \caption{Formation energy for $Ni_2XMn$ (X=Co,Cr,Fe) as a function of volume.} \label{fig:Modelo-Lineal}
\end{figure*}

In figure \ref{fig:Modelo-Lineal} we summarize the formation energy, related with pure elements energy in bcc structure, following the equation $ E_{form}[Ni_2MnX] = E_{str}[Ni_2MnX] - \left [ 2 E_{bcc} Ni + E_{bcc} Ni +E_{bcc} X \right ]$, as a function of the isotropic expansion for the three alloys studied. For instance, in the subfigure \ref{fig:Ni2FeMn} we show the behavior of the structures $L2_1$ (triangles) and $f\bar{4}3m$ (diamonds) in ferromagnetic (blue color) and antiferromagnetic (red color) order, where it can be seen that both structures $L2_1$ are energetically favorable compared to $f\bar{4}3m$; regarding magnetization, the energy difference between the magnetic configurations for $L2_1$ in the equilibrium position is small, favoring the AFM structure, and it is observed that there is an important difference in the equilibrium parameter. In relation with the FM and AFM magnetic configurations for $f\bar{4}3m$, the energy difference becomes more important, also showing a lower energy for the AFM magnetization. In table \ref{tab:Resumen} we show the values of the equilibrium structure obtained by means a Birch-Murnaghan equation fit. Bulk modulus values for this alloy is around 180 GPa, with the exception of $f\bar{4}3m-FM$, which has a considerably lower value.

In the figure \ref{fig:Ni2CoMn}, the formation energy values for the $Ni_2CoMn$ alloy are observed. Unlike the previous case, it is observed that both FM magnetizations are stable compared to AFM; being the structure $f\bar{4}3m-FM$ the equilibrium structure, in addition to being the only one that has a negative formation energy. In addition, the energy difference between the structure $L2_1$ and $f\bar{4}3m$ in AFM magnetization is interesting. The equilibrium parameters strongly depend on both the magnetization and the structure. The same occurs with the Bulk modulus value and its derivative, while the $L2_1-FM$ structure has a modulus of 226.00 GPa, the $f\bar{4}3m-FM$ structure has a modulus of $20\% $ minor.

{\small
\begin{table*}[thb]
    \centering
    \caption{Energy of formation, lattice parameter, bulk modulus, their derivative, magnetic moments per atomic specie and Fermi energy for the studied alloys. Equilibrium structures are marked with arrow.}
\begin{tabular}{|c|c| c|c |c |c | c |c | c | c  |c | }
\hline 
Alloy & Structure & $E_{f}$ [meV] & $a_0$[a.u.] & $  B $[GPa] & $B' $ & $\mu_{Ni_1}$  &  $\mu_{Ni_2}$ &   $\mu_{Mn}$ & $\mu_{X}$ & $E_F$[eV] \\ \hline
$Ni_2MnFe$ & & & & & & & & & &\\ 
 & $L2_1-FM$ &  -3.36 &  10.734 &  184.50 & 4.577 & 0.8267 & 0.8267 & 3.2483  &  2.8399   &  17.9452 \\ 
  $\longrightarrow$& $L2_1-AFM$ &  -4.32 &  10.652 &  180.38 & 4.555 & 0.0570 & 0.0570 & 3.0673 &  -2.7149   &  18.2999 \\
 & $f\bar{4}3m-FM$ &  30.88 &  10.596 &  160.09 & 3.923  & 0.5752 & 0.7051 & 2.7070  &  2.4757   & 18.3850 \\ 
 & $f\bar{4}3m-AFM$ &  2.18 &  10.557 &  189.35 & 5.020   & 0.4017 & -0.4642 & 2.5229  & -2.3087  & 18.8613\\
\hline
$Ni_2MnCo$ & & & & & & & & & &\\ 
 & $L2_1-FM$ &  33.74 &  10.617 &  226.00 & 2.119  & 0.7885 & 0.7885 & 3.1731  & 1.7551 & 18.2124 \\ 
 & $L2_1-AFM$ &  29.11 &  10.559 &  189.16 & 4.684  & 0.2264 & 0.2264  &  2.8964 & -1.6348 & 18.5213  \\
  $\longrightarrow$  & $f\bar{4}3m-FM$ &  -4.47 &  10.573 &  177.15 & 3.638  &  0.6794 & 0.6293 &  2.8945 & 1.7354 & 18.2998\\ 
 & $f\bar{4}3m-AFM$ &  34.62 &  10.465 &  191.11 & 6.341  &  0.3204 & -0.4432 & 2.6050 & -0.3938 & 18.9432 \\
\hline
$Ni_2MnCr$ & & & & & & & & & & \\ 
 & $L2_1-FM$ &  133.88 & 10.817 &  141.1 & 2.827 &  0.6711 &  0.6711 &  3.2502 & 2.3276 &  17.7548  \\ 
 & $L2_1-AFM$ &  86.34 & 10.765 &  167.5 & 4.633  & 0.0189 & 0.0189 & 3.1572 & -2.3055 &  17.9238 \\
  $\longrightarrow$  & $f\bar{4}3m-FM$ &  -2.85 & 10.587 &  183.41 & 5.401  & 0.5747 & -0.2396 & 2.2598  & -1.6752  & 18.8633 \\ 
 & $f\bar{4}3m-AFM$ &  61.77 &  10.453 &  266.9 & 3.608  & -0.0481& 0.1672 &   0.3730 & -0.1086    & 19.8133 \\
\hline
\end{tabular}
\label{tab:Resumen}
\end{table*}

}

Finally, in figure \ref{fig:Ni2CrMn}, we show formation energy values for the alloy $Ni_2CrMn$, where the equilibrium structure is $f\bar{4}3m-FM$. For this alloy, a trend is not observed as in the previous cases where they are grouped by type of structure or magnetization. In fact, also the equilibrium lattice parameters differ a lot from each other, being the smallest the one corresponding to $f\bar{4}3m-AFM$; in turn, it has a much larger bulk modulus value than the rest of the structures. 

Magnetic moments per atomic specie and Fermi energy are displayed for equilibrium lattice parameter in every structure.  As expected, magnetic moments for $Ni_1$ and $Ni_2$ in $L2_1$ order have the same value. Instead, for $f\bar{4}3m $ order this behaviour is not conserved. The magnetic moment values of the manganese atoms are higher than for the remaining atoms and play a major role in the final magnetization. Magnetic moment of element $X$ establish the character ferro- or antiferromagnetic of the structure. Nickel's magnetic moment is one order below and has the same sign as the predominant element (Mn or X) that is its first neighbor. Fermi energy is, in all the cases, higher for AFM order then FM order. In the same way, it is also higher for $f\bar{4}3m$ than for $L2_1$. In the following, we will work only with the stable structures marked with a arrow in the table \ref{tab:Resumen}.

\begin{figure}[th]
  \begin{center}
    \includegraphics[width=0.45\textwidth]{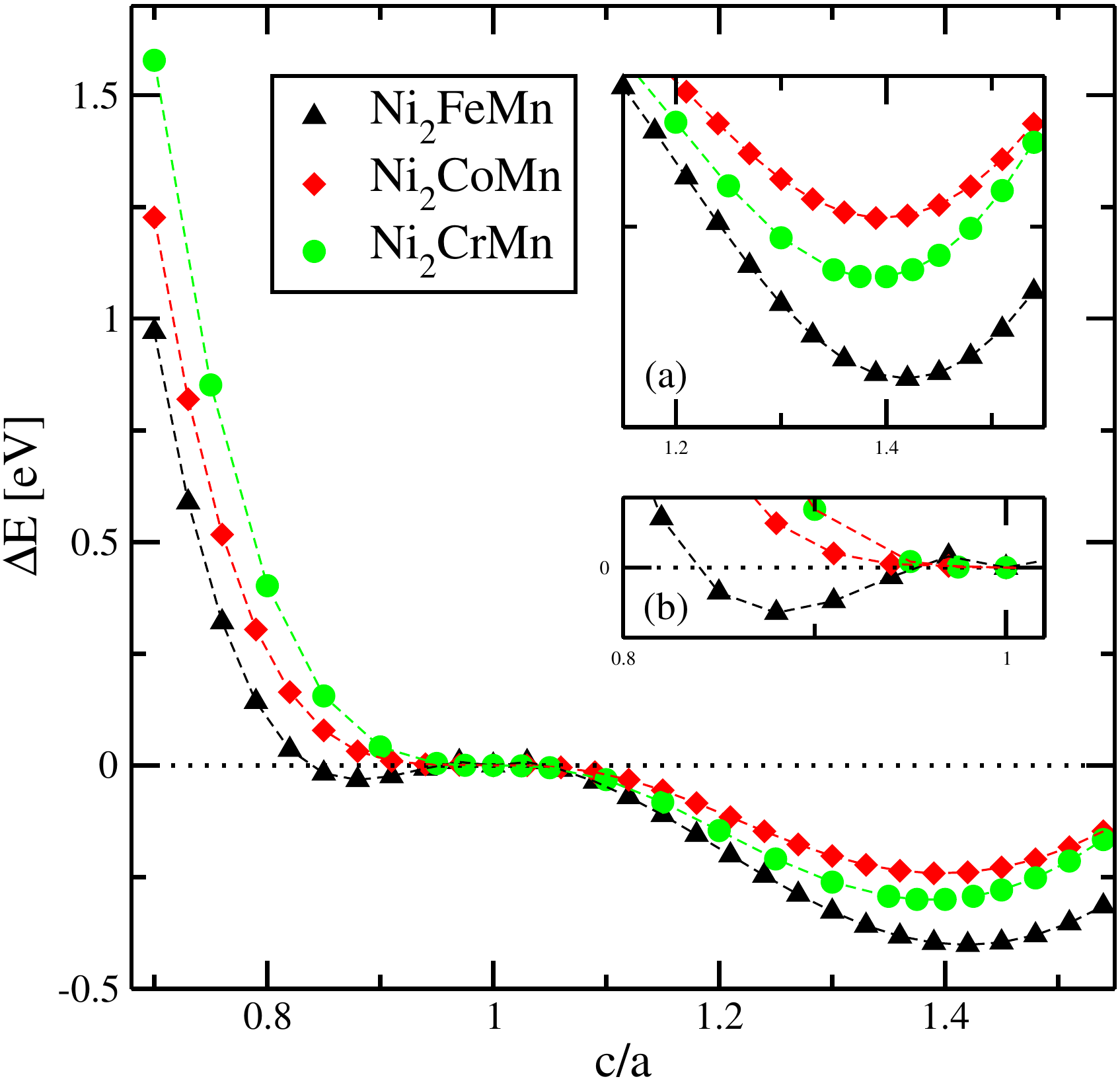}
     \caption{ Difference of energy as a function of tetragonal distortion for the alloys studied. In insets, the focus is on different regions: in inset (a) region around $c/a \sim 0.9$, inset (b) around $c/a \sim 1.2-1.5$. }
     \label{fig:Bain}
 \end{center}
\end{figure}

Tetragonal distortion was carried out using a volume-conserving Bain path that consists which consists of modifying the $c/a$ ratio of the z-axis with the x- and y-axes\cite{ozolins2009first}. In this study, we use the relation of the $c/a$ is equal to the unity for bcc-type  and for fcc-type is $\sqrt 2$. A $c/a$ ratio between $0.70$ and $1.50$ has been studied for every alloy. We show the behavior of difference of energy as a function of $c/a$ distortion for three alloys in Fig. \ref{fig:Bain}. As a common result, a region of local minima is observed for $c/a = 1$ and global minima for values $c/a $ near to $ \sqrt 2$, which were amplified in inset (a) of the figure. For the $Ni_2CrMn$ alloy we find that the minimum corresponds to $c/a \approx 1.38$ and presents an energy difference with respect to the structure derived from the bcc of $\Delta E \approx 0.24 eV $ by unitary formula and, in the case of $Ni_2CoMn$ we find $c/a \approx 1.39$ and $\Delta E \approx 0.31 eV $. The $Ni_2FeMn$ alloy, for its part, presents a global minimum at $c/a \approx 1.41$ with an energy difference of $\Delta E \approx 0.40 eV $, which can be seen, and a local minimum at $c /a \approx 0.88$ and $\Delta E \approx 0.09 eV $, magnified in inset (b) of the figure for detailed view. This minimum could be compatible with the appearance of modulate stacking faults, as has been seen in other Ni-Mn alloys, i.e. \cite{niemann2012role,gruner2018modulations}.

\begin{figure}[htb]
  \begin{center}
    \includegraphics[width=0.5\textwidth]{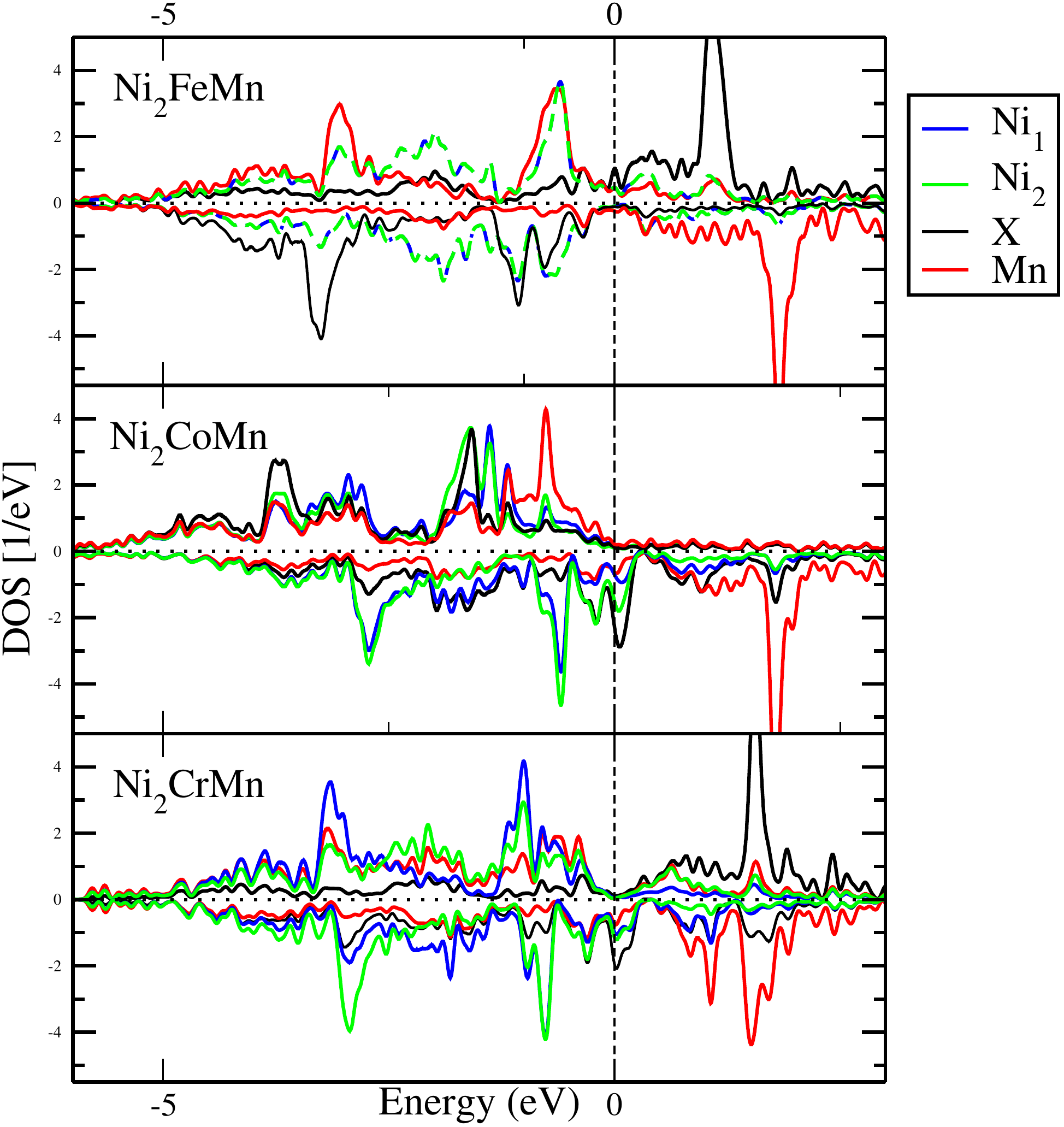}
     \caption{Density of electronic states for $Ni_2FeMn$, $Ni_2CoMn$ and $Ni_2CrMn$.}
     \label{fig:Bandas}
 \end{center}
\end{figure}

More detailed information for the understanding of the magnetic behavior can be obtained from the analysis of the density of electronic states. The Fig. \ref{fig:Bandas} we show partial DOS, according to the atomic species, for the three alloys studied. The energy axis has been shifted so that the zero of the x-axis corresponds to the Fermi energy. Positive (negative) values of electronic density represent spin-up (spin-down). For all cases, spin-up correspond to majority spin contribution. The subfigure above shows the partial DOS, separated by atomic species, obtained for $Ni_2FeMn$. The superposition of the DOS for the atoms $Ni_1$ and $Ni_2$ is clearly observed, corresponding to the symmetry that the structure $L2_1$ presents for the majority element. Below the Fermi energy, the DOS for the manganese atom condenses in the spin-up region while the opposite happens for the iron atom. Consequently, above the Fermi energy there are two peaks of opposite spin, corresponding to the iron and manganese atoms with different energies. The density distribution for nickel atoms is more symmetric. Near the Fermi energy the spin-down states vanish or have very small values, which indicates near half-metallic behavior of the alloy \cite{galanakis2002slater}.

In the central subfigure we show the DOS for $Ni_2CoMn$, where the peaks of the $Ni_1$ and $Ni_2$ atoms differ from each other, as corresponds to a $f\bar{4}3m$ structure. In this case, the partial densities for the manganese and cobalt atoms are grouped in the spin-up region, causing an asymmetry in the regions corresponding to ferromagnetic behavior. The analysis at energies near to the Fermi energy shows half-metallic behavior. Above this energy there is an important peak corresponding to a spin-down state of the manganese atom.

Finally, in the subfigure below we plot the partial DOS for the $Ni_2CrMn$ alloy. Unlike the previous cases, a dominant role of the manganese atom is not seen with respect to the partial densities of the remaining atoms. Likewise, the contribution of the chromium atom for energies close to the Fermi energy is much lower than that observed in the previous cases of iron and cobalt atoms; in this case, nickel atoms have a greater preponderance. Due to the slight difference between spin-up and spin-down states for the different atomic species, the alloy is of the ferrimagnetic type. As in the previous case, half-metallicity is also observed.

\section{Conclusions}
\label{sc:CC} 

In this article, spin-polarized ab-initio calculations have been carried out on alloys with composition $Ni_2XMn$, where X is a d-transition metal, in Heusler and inverse Heusler type structures, for different initial magnetizations. The stability of the magnetic structures was established and various parameters of interest extracted from the equation of state were obtained by means of the equation of state. As a main result, it is observed that the magnetic interactions are governed by the element X and the manganese atom. On other hand, Nickel's contribution to magnetization is considerably smaller.

When tetragonal distortion is considered, all the compositions present a local minimum at $c/a=1$ and an global minimum near $c/a \sim \sqrt{2}$. Furthermore, the $Ni_2FeMn$ alloy has an additional local minimum around $c/a \sim 0.88$, where it is possible to find interesting modulate stacking faults. Difference of energies are reported in the text.

An analysis of the electronic DOS in this type of alloys was performed to understand the underlying magnetic phenomena. The different peaks for the partial DOS were identified depending on the atomic species involved and the asymmetries that leading magnetic effects have been highlighted. Like other Heusler alloys composed only of all d-transition metals, these alloys show half-metallic behavior, which makes them of interest for technological and engineering applications. 

\section*{Acknowledgments}
This work was supported by ANPCyT (PICT 2017-4062) and CONICET (PIP  11220200102859CO). Computational facilities had been provided by the Centro de Computación de Alto Desempeño Tandil (CCADT), Facultad de Ciencias Exactas, UNCPBA.  Author acknowledges postdoctoral fellowship of CONICET-Argentina. Figure \ref{fig:Estructuras} was made using VESTA software \cite{momma2008vesta}.

 {\small
 
 \printbibliography}
 
\end{document}